\documentstyle[preprint,aps,epsfig]{revtex}
\begin{document}

\tighten
\draft
\preprint{
\vbox{
\hbox{December 2000}
\hbox{Tashkent}
}}
\newcommand{\re}[1]{(\ref{#1})}
\newcommand{\lab}[1]{\label{#1}}
\newcommand{\ci}[1]{\cite{#1}}
\renewcommand{\baselinestretch}{1.25}
\newcommand{\bfr}{\begin{flushright}}
\newcommand{\bfl}{\begin{flushleft}}
\newcommand{\efl}{\end{flushleft}}
\newcommand{\efr}{\end{flushright}}
\newcommand{\bc}{\begin{center}}
\newcommand{\ec}{\end{center}}
\newcommand{\be}{\begin{equation}}
\newcommand{\ee}{\end{equation}}
\newcommand{\bea}{\begin{eqnarray}}
\newcommand{\eea}{\end{eqnarray}}
\newcommand{\ba}{\begin{array}}
\newcommand{\ea}{\end{array}}
\newcommand{\edc}{\end{document}}
\newcommand{\ul}{\underline}
\newcommand{\ri}{\rightarrow\infty}
\newcommand{\li}{\leftarrow\infty}
\newcommand{\ra}{\rightarrow}
\newcommand{\la}{\leftarrow}
\newcommand{\ds}{\displaystyle}
\newcommand{\dsf}{\displaystyle\frac}
\newcommand{\dt}{\Delta{t}}
\newcommand{\il}{\int\limits}
\newcommand{\pal}{\partial}
\newcommand{\xxx}{{\it{X}}}
\newcommand{\bone}{{\bf 1}}
\title{
Approximate formula for the ground state energy\\ of anyons 
in $2D$ parabolic well}

\author{B. Abdullaev$^1$, M. Musakhanov$^1$, A. Nakamura$^{2}$ }
\address{$^1$
 Theoretical  Physics Dept, Uzbekistan National University,\\
 Tashkent 700174, Uzbekistan\\
e-mail: bah$\_$abd@iaph.silk.org, yousuf@iaph.silk.org}
\address{$^2$
RIISE, Hiroshima University, Japan}

\maketitle

\begin{abstract}

We determine approximate formula for the ground state energy of anyons in
$2D$ parabolic well which is valid for the arbitrary anyonic factor $\nu$
and number of particles $N$ in the system. We assume that centre of mass
motion energy is not excluded from the energy of the system. 
Formula for  ground
state energy calculated by variational principle contains logarithmic
divergence at small distances between two anyons which is regularized by
cut-off parameter. 
By equating this variational formula to the analogous
formula of Wu near
bosonic limit ($\nu\sim 0$)we determine the value of the cut-off 
and thus derive the approximate formula for the ground state energy
for the any $\nu$ and $N$. 
We checked this formula at $\nu=1$, when anyons become fermions,
for the systems containing two to thirty particles.
We find that our approximate formula has an accuracy within $6\%$. 
It turns out, at the big number $N$
limit the ground state energy has square root dependence on factor $\nu$.

\end{abstract}

\section{Introduction}

Anyons - two-dimensional particles obeying fractional statistics are one
of possible quasiparticle excitations in fractional quantum Hall effect
and High $T_c$ superconductors \ci{1,2,3,4,5}.

The determination of the energy spectrum of anyons in $2D$ parabolic well
is actual and interesting task of quantum mechanics. The analytical
calculation of the energy spectrum of the two anyons in $2D$ parabolic
well was performed in \ci{6,7}. The energy spectrum of systems containing
more than two anyons in the same external field can not be obtained
analytically. Wu \ci{8} generalized the class of exact two anyon solution
into three anyons containing system and found it's energy spectrum.
There are the numerical calculations of the lowest part of the energy
spectrum of three \ci{9,10} and four \ci{11} anyons in $2D$ parabolic well.

There is separate quantization  of centre  mass motion  energy and
relative motion energy to each other of anyons in $2D$ parabolic well
\ci{7,8}.  The ground state of relative motion energy of the systems of two,
three and four anyons can be linearly interpolated between bosonic and
fermionic spectra \ci{6,7,8,9,10,11} when anyonic factor $\nu$ changes from 0
to 1.

Linear dependence of full ground state energy ( without separation of cenre
mass motion energy ) from anyonic factor $\nu$ exists for arbitrary number
of anyons only near the bosonic limit of spectra \ci{8}, i.e.
at $\nu\rightarrow0$. Beginning with the case of three anyons it deviates from
linear character near the fermionic limit of spectra \ci{8}, i.e. at
$\nu\rightarrow 1$.

By variational calculation we determine approximate formula for the
ground state energy of anyons in
$2D$ parabolic well which is valid for the arbitrary anyonic factor $\nu$
and number of particles $N$ in system. We assume that centre of mass
motion energy is not excluded from energy of system and anyons represented
by bosonic variational wave function. As in Laughlin's variational calculation
of infinite  system of free anyons \ci{12}, our
formula for the  ground
state energy calculated by variational principle contains logarithmic
divergence at small distances between two anyons regularized by cut-off.
 We equate our variational ground state energy formula to the
same formula of Wu \ci{8} for anyons in $2D$ parabolic well near the
bosonic limit on the factor $\nu$ to determine the cut-off 
and thus derive the approximate formula for the ground state energy
at the any factor $\nu$ and number $N$. 
We check this formula at $\nu=1$, when anyons become fermions,
for the systems containing two to thirty particles.
We find that our approximate formula has an accuracy within $6\%$. 
It turns out, at the big number $N$
limit the ground state energy has square root dependence on factor $\nu$.

The following section 2 describes the hamiltonian of
the system of anyons in $2D$ parabolic well and the choice 
of trial wave function
for the variational calculation and physical argument for this choice.
Section 3 contains results of variational calculation.

\section{ Hamiltonian and wave function of system}

The Hamiltonian of $N$ anyons system in $2D$ parabolic well has a form:
\be
\hat H=\dsf{1}{2M}\ds\sum_{i=1}^N(\vec p_i+\vec A(\vec r_i))^2+
\dsf{M\omega_o^2}{2}\ds\sum_{i=1}^N\vec r_i^2.
\lab{1}
\ee
Here $M$ - mass of particle,  $\vec p=-i\hbar \vec
\nabla$, where  $\vec \nabla=\vec i \dsf{\pal}{\pal x}+
\vec j \dsf{\pal}{\pal y}$, $\omega_o$ - eigen frequency of oscillatory motion
of free
particles in $2D$ parabolic well and $\vec r_i$ - radius vector of $i$ - th
particle number.

Vector potential for anyons $\vec A(\vec r_i)$ \ci{8,12} in \re{1} is
\be
\vec A(\vec r_i)=\hbar\nu\ds\sum_{j>i}^N\dsf{\vec z \times\vec r_{ij}}
{|\vec r_{ij}|^2}.
\lab{2}
\ee
Here $\vec z$ is unit vector perpendicular to $2D$ plane where situated
system,$\vec r_{ij}=\vec r_i-\vec r_j$  and anyonic factor $\nu$
characterizes the form of fractional statistics ( spin of the anyon ):
it can be changed from $\nu=0$ - noninteracting bosons to $\nu=1$
 - free fermions.

In the variational consideration  the ground state energy of system
determined as minimum of energy
\be
E=\dsf{\int \Psi_T^*(\vec R)\hat H \Psi_T(\vec R) d\vec R}{\int \Psi_T^*(\vec R)
\Psi_T(\vec R) d\vec R}.
\lab{3}
\ee
Here $\hat H$ - the Hamiltonian, $\Psi_T(\vec R)$ - trial wave function
of system that has a variational parameter and $\vec R$ - the coordinates
of all particles. Minimum energy of $E$ \re{3} is reached  by variation of
variational parameter of trial wave function.

We suppose that trial wave function is provided by mean field approximation
of Fetter,Hanna and Laughlin \ci{17}. In this  approximation particles move
in the mean
field $\overline{\vec {A}}$:
\be
\overline {\vec{A}}(\vec r) =
\dsf{1}{2}\vec B\times\vec r
\lab{4}
\ee
induced by the average density $\rho$ of particles in the system where
$\vec B=2\pi\rho\hbar\nu\vec z$, $\rho=\dsf{1}{\pi r_o^2}$ and
$r_o$ - mean distance between particles.
According magnetic field $\vec B$ one can introduce corresponding magnetic length
$a_H=(\hbar/B)^{1/2}$
and cyclotron frequency $\omega_c=B/M$.
Energetic
spectrum of Hamiltonian \re{1} with $\overline {\vec A}$ and $2D$ parabolic well
has the same energetic spectrum structure as a free particle in $2D$
parabolic well \ci{18}.

Thus assuming a bosonic representation of anyons
\be
\Psi_T (\vec R)=\prod_{i=1}^N\Psi_T (\vec r_i)
\lab{5}
\ee
for the ground state of anyons in $2D$ parabolic well one can bring
wave function $\Psi_T (\vec r_i)$ in the following form:
\be
\Psi_T(\vec r_i)=C\exp\left(-(\alpha+\nu)\dsf{(x_i^2+y_i^2)}{2r_o^2}\right).
\lab{6}
\ee
Here C - normalization constant, $\alpha$ variational parameter and there
assumed that mean distance between particles $r_o$ equals of
$R_o=(\hbar/M\omega_o)^{1/2}$ - raduis localization for the wave function
in the ground  state of $2D$ oscillator. This equality follows from fact that
interaction between two anyons at any not zero anyonic factor $\nu$
has repulsive character and relation $S_n=n\pi R_o^2$ where $S_n$ - area
that occupies by free particle at $n$ - th energetic level of $2D$ oscillator.
Last relation for the area $S_n$ one can easily get by using a virial
theorem for $2D$ oscillator.

It is convenient to express all energetic quantities by energetic quanta
of $2D$ oscillator $\hbar\omega_o$ and length quantities by $r_o$.

Normalized trial wave function of system has a form:
\be
\Psi_T(\vec R)=\left(\dsf{\alpha+\nu}{\pi}\right)^{N/2}\prod_{i=1}^N
\exp\left(-(\alpha+\nu)\dsf{(x_i^2+y_i^2)}{2}\right).
\lab{7}
\ee

\section{Variational consideration. Result of calculation.}

At the calculation of variational energy $E$ \re{3} it is convenient to deal
with quantity  $E_L(\vec R)=\Psi_T^{-1}(\vec R)\hat H \Psi_T(\vec R)$.

As trial wave function $\Psi_T(\vec R)$ is not true of the Hamiltonian \re{1},
so
\be
E_L(\vec R)=Re E_L(\vec R)+i Im E_L(\vec R).
\lab{8}
\ee

By the acting of the Hamiltonian $\hat H$ \re{1} on $\Psi_T(\vec R)$ \re{7}
we find
\be
Re E_L(\vec R)=\ds\sum_{i=1}^N[\alpha+\nu+\dsf{x_i^2+y_i^2}{2}-
\dsf{(\alpha+\nu)^2}{2}(x_i^2+y_i^2)+\dsf{\nu^2}{2}(\vec A(\vec r_i))^2],
\lab{9}
\ee
and
\be
Im E_L(\vec R)=-\nu (\alpha+\nu)\ds\sum_{i=1}^N(\vec A(\vec r_i)\vec r_i).
\lab{10}
\ee
In the expression \re{10} we have a scalar product of vector $\vec A(\vec r_i)$
and $\vec r_i$.

A variational energy $E$ \re{3} expressed by $E_L(\vec R)$ is
\be
E=\int \Psi_T(\vec R)\ E_L(\vec R)\Psi_T(\vec R) d\vec R.
\lab{11}
\ee

Direct calculation gives
\be
\int \Psi_T(\vec R)\ Im E_L(\vec R)\Psi_T(\vec R) d\vec R=0.
\lab{12}
\ee
So, on energy $E$ contributes only $Re E_L(\vec R)$.

Integrals \re{11} with $\Psi_T(\vec R)$ \re{7} of first three terms in square
brackets \re{9} are calculating elementary. Problems appear when we are
calculating  integrals with term proportional of $(\vec A(\vec r_i))^2$.
This term gives two kind integrals: integrals like
$\int \Psi_T(\vec R)\ \dsf{1}{|\vec r_{ij}|^2}\Psi_T(\vec R) d\vec R$
where $i\ne j$ number of which are $N(N-1)$ and integrals like
$\int \Psi_T(\vec R)\ \dsf{1}{|\vec r_{ij}|} \dsf{1}{|\vec r_{ik}|}
\Psi_T(\vec R) d\vec R$
where $i\ne j$, $i\ne k$ and $j\ne k$  number of which are $N(N-1)(N-2)$.

By taking account of integrals ( see Gradshtein and Ryzik \ci{19}):
\be
\int_{0}^{\infty} E_i(ax)e^{-\mu x}dx=-\dsf{1}{\mu}\ln
\left(\dsf{\mu}{a}-1\right),
\lab{13}
\ee
where constants $a>0$, $Re \mu>0$ and $\mu>a$ and
\be
\int_{0}^{\infty} E_i(-\beta x) e^{-\mu x}dx=\dsf{1}{\mu}\ln
\left(\dsf{\mu}{\beta}+1\right),
\lab{14}
\ee
where constants $Re (\beta+\mu)\geq 0, \mu>0$ and $E_i(\gamma y)=
-vp \int_{-y}^{\infty} e^{-\gamma z}dz/z$ - the exponential integral
where $\gamma>0$ and
$vp\int$ indicates integral in main quantity mean
one can get
\be
\int \Psi_T(\vec R) \dsf{1}{|\vec r_{ij}|^2}\Psi_T(\vec R) d\vec R=
(\alpha+\nu)\ln\left(\dsf{2}{\delta}\right),
\lab{15}
\ee
where cut-off  parameter $\delta$ regularize the integral \re{15}.

And an other hand the calculation gives
\be
\int \Psi_T(\vec R)\ \dsf{1}{|\vec r_{ij}|} \dsf{1}{|\vec r_{ik}|}
\Psi_T(\vec R) d\vec R=(\alpha+\nu)A,
\lab{16}
\ee
where numerical value $A\approx 1$.

So, term by term averaged quantity $Re E_L(\vec R)$ \re{9} gives expression
for energy
\be
E=N(\alpha+\nu)+\dsf{N}{2(\alpha+\nu)}-\dsf{N(\alpha+\nu)}{2}+\\
\dsf{1}{2} \nu^2 N(N-1)(\alpha+\nu)[\ln\left(\dsf{2}{\delta}\right)-
(N-2)] .
\lab{17}
\ee
The extremum condition $\dsf{dE}{d(\alpha+\nu)}=0$ gives a parameter
\be
(\alpha+\nu)=[1+ \nu^2 (N-1)[\ln\left(\dsf{2}{\delta}\right)-
(N-2)]]^{-1/2}
\lab{18}
\ee
at which minimum energy of $E$ is
\be
E_v=N[1+ \nu^2 (N-1)[\ln\left(\dsf{2}{\delta}\right)-
(N-2)]]^{1/2} .
\lab{19}
\ee
As established Wu \ci{8} near the bosonic limit of ground state
energy of anyons in $2D$ parabolic well on the factor $\nu$, i.e.
$\nu \rightarrow 0$, ground state energy when centre mass motion
not fixed has a following expression ( here and below in this section
we return into normal units of energy and length):
\be
E \approx \hbar\omega_o [N+N(N-1)\nu/2]
\lab{20}
\ee
for the arbitrary number of particles $N$ in system.
So, at $\nu \rightarrow 0$ we expand expression for the energy $E_v$
\re{19} on the powers $\nu^2$ and by equating term proportional to $\nu^2$
of expansion to second term in square brackets of \re{20}, we find that
\be
\delta=2\exp\left( -\dsf{1+\nu(N-2)}{\nu}\right)r_o
\lab{21}
\ee
and finally analytical expression for the variational ground state energy is:
\be
E_v=\hbar\omega_o N[1+\nu(N-1)]^{1/2}.
\lab{22}
\ee
At the Table 1 we bring exact energies for fermions in $2D$ parabolic
well constructed by filling lowest energetic levels of $2D$ oscillator
and energies calculated by formula \re{22}. From this table one can see
that deviation of variational energy $E_v$ from exact energy
$E_{EXACT}$ is no more than $6\%$.

\begin{table}
\caption{The ground state energy of $N$ fermions
in $2D$ parabolic well.\\
$E_{EXACT}$ - the energies  
given by exact analytical formula
from \protect\ci{18}
(in $\hbar\omega_o$ units),\\
$E_v$ - the energies
(in $\hbar\omega_o$ units) calculated by formula \protect\re{22}. }
\begin {center}
\begin{tabular}
{|c|c|c|c|}
\hline
{$N$}&{$E_{EXACT}$}&{$E_v$}
&{$(E_v - E_{EXACT})/E_{EXACT}$ in $\%$}\\ \hline
{2}&{3}&{2.83}&{6}\\ \hline
{3}&{5}&{5.20}&{4}\\ \hline
{4}&{8}&{8.00}&{0}\\ \hline
{5}&{11}&{11.18}&{2}\\ \hline
{6}&{14}&{14.70}&{5}\\ \hline
{7}&{18}&{18.52}&{3}\\ \hline
{8}&{22}&{22.63}&{3}\\ \hline
{9}&{26}&{27.00}&{4}\\ \hline
{10}&{30}&{31.62}&{5}\\ \hline
{11}&{35}&{36.48}&{4}\\ \hline
{12}&{40}&{41.57}&{4}\\ \hline
{13}&{45}&{46.87}&{4}\\ \hline
{14}&{50}&{52.38}&{5}\\ \hline
{15}&{55}&{58.09}&{6}\\ \hline
{16}&{61}&{64.00}&{5}\\ \hline
{17}&{67}&{70.09}&{5}\\ \hline
{18}&{73}&{76.37}&{5}\\ \hline
{19}&{79}&{82.82}&{5}\\ \hline
{20}&{85}&{89.44}&{5}\\ \hline
{21}&{91}&{96.23}&{6}\\ \hline
{22}&{98}&{103.19}&{5}\\ \hline
{23}&{105}&{110.30}&{5}\\ \hline
{24}&{112}&{117.58}&{5}\\ \hline
{25}&{119}&{125.00}&{5}\\ \hline
{26}&{126}&{132.57}&{5}\\ \hline
{27}&{133}&{140.30}&{5}\\ \hline
{28}&{140}&{148.16}&{6}\\ \hline
{29}&{148}&{156.17}&{6}\\ \hline
{30}&{156}&{164.32}&{5}

\end{tabular}
\end{center}
\end{table}


\begin{thebibliography}{9}

\bibitem{1}
F. Wilczek, Fractional Statistics and Anyon Superconductivity,
World Scientific, Singapore, 1990.


\bibitem{2}
S. Forte, {\it Rev.Mod.Phys.,}{\bf 64},193 (1992).

\bibitem{3}
R. Iengo and K. Lechner, {\it Phys.Rep.,}{\bf 213}, 179 (1992).

\bibitem{4}
A. Lerda, Anyons, Lecture Notes in Physics m14,
Springer-Verlag, Berlin-Heidelberg, 1992.

\bibitem{5}
M. Stone, editor, Quantum Hall Effect, World Scientific, Singapore, 1992.

\bibitem{6}
J.M. Leinaas and J. Myrheim, {\it Nuovo Cimento, }{\bf B37}, 1 (1977).

\bibitem{7}
F. Wilczek, {\it Phys.Rev.Lett.,}{\bf 48}, 1144 (1982).

\bibitem{8}
Y.-S. Wu, {\it Phys.Rev.Let.,}{\bf 53}, 111 (1984), Erratum {\it ibid}
{\bf 53}, 1028 (1984).

\bibitem{9}
M. Sporre, J.J.M. Verbaarschot and I. Zahed,
{\it Phys.Rev.Lett.,}{\bf 67}, 1813 (1991).

\bibitem{10}
M.V.N. Murthy, J. Law, M. Brack and R.K. Bhaduri,
{\it Phys.Rev.Lett.,}{\bf 67}, 1817 (1991).

\bibitem{11}
M. Sporre, J.J.M. Verbaarschot and I. Zahed,
{\it Phys.Rev.}{\bf B46}, 5738 (1992).

\bibitem{12}
R.B. Laughlin, {\it Phys.Rev.Lett.,}{\bf 60}, 2677 (1988).


\bibitem{17}
A.L. Fetter, C.B. Hanna and R.B. Laughlin,
{\it Phys.Rev.}{\bf B39}, 9679 (1989).

\bibitem{18}
V.M. Galicki, B.M. Karnakov and V.I.Kogan, "The problems on the
quantum mechanics"(in Russian), Moscow, "Nauka", 1981, 648.

\bibitem{19}
I.S. Gradshtein and I.M. Ryzik, Tables of integrals, sums, expansions
and products (in Russian), Moscow, Fizmatgiz, 1100, 1962.

\end{thebibliography}
 \end{document}